# The Monte Carlo and Fractional Kinetics approaches to the underground anomalous subdiffusion of contaminants

M. Marseguerra [1], A. Zoia [2]

*Department of Nuclear Engineering, Polytechnic of Milan, Via Ponzio 34/3, 20133 Milan, Italy*

## Abstract

It is nowadays recognized that the experimental evidences of many transport phenomena must be interpreted in the framework of the so-called anomalous diffusion: this is the case e.g. of the spread of contaminant particles in porous media, whose mean squared displacement (MSD) has been experimentally reported to grow in time as $t^\alpha$ (with $0<\alpha<2$). This is in deep contrast with the linear increase which characterizes the standard Fickian diffusive processes. Anomalous transport is currently tackled within the Fractional Diffusion Equation (FDE) analytical model, which has blown new life into the concept of fractional derivatives, dormant since Leibniz's first discovery, about 300 years ago.

In this paper we focus on subdiffusion, i.e. the case $\alpha<1$, which provides a suitable explanation of the observed non-Fickian persistence of the contaminant particles near the source in various transport experiences in heterogeneous media. Unfortunately, the FDE approach requires several approximations to overcome its analytical complexities. To evaluate the relevance of these approximations, we propound the use of the Monte Carlo simulation as a suitable reference for the analytical FDE results. This comparison shows that the FDE results are less and less reliable as $\alpha$ becomes closer to *1*. The approach herein proposed can be easily applied to various fields of science where anomalous diffusion phenomena may occur, from physics and chemistry to biology and finance.

## 1. Introduction

About eighty years ago, Richardson, investigating the diffusion of a pair of fluid elements in fully developed turbulence, observed the existence of transport phenomena characterized by a mean square displacement (MSD) proportional not to the time *t* (as in the usual diffusion equation established by Fick in 1855 (Fick 1855)), but instead to the third power of *t* (Richardson 1926). This observation opened in the late 1960 the route to the investigations on a completely new kind of diffusion, in the framework of the Fractional Kinetics: nowadays, transport phenomena characterized by a MSD proportional to $t^\alpha$, with $\alpha \neq 1$, are called anomalous diffusion processes, even if they are so ubiquitous that it might be reasonably said that they represent the norm. If α < 1 the diffusion is slower than in the case of standard diffusion and it is called subdiffusion; if α > 1 the diffusion is faster and it is called superdiffusion. Since the Richardson's work, a lot of experimental evidences concerning the existence and the exploitation of anomalous diffusions have been obtained in quite different fields, such as physics, chemistry, biology, economics, geology. Most of the theoretical interpretations have been established within the Fractional Kinetics approach, which generalizes the Fick's second law, the Fokker-Planck equation and the Montroll and Weiss (Montroll, Weiss 1965) framework of the continuous time random walks (CTRW). Exhaustive references may be found e.g. in the fundamental papers by Metzler and Klafter (Metzler, Klafter 2000; Metzler, Klafter 2004). Just to give an idea of the possible applications, we explicitly mention only some important results, e.g. the subdiffusive charge transport in amorphous semiconductors (Scher, Montroll 1975), the anomalous diffusion properties of heat channels (Denisov et al. 2003), the chemical reaction processes in high-*k* dielectric films (de Almeida, Baumvol 2003), the motion of DNA-binding proteins along the DNA structure (Slutsky et al. 2003), the waiting time between two financial transactions (Mainardi et al. 2000), the contaminant transport in geological formations (Kosakowski 2004).

Here we focus on this last subject: the underground migration of contaminants driven by the subsurface water flow, where the medium heterogeneities over many scales and the trapping of the contaminant particles in the solid environment support the occurrence of subdiffusion phenomena. The importance of this problem is well recognized particularly in view of the possibility of leakage and successive migration of long-term radioactive or toxic species from underground repositories. We tackle this problem within the Monte Carlo approach, whose results are compared to the analytical ones of the fractional kinetics.

---

[1] *Corresponding Author. Tel: +39 02 2399 6355 . Fax: +39 02 2399 6309*
*Email address: marzio.marseguerra@polimi.it (Marzio Marseguerra)*

[2] *Email address: andrea.zoia@polimi.it (Andrea Zoia)*

In Section 2 we summarize the Fractional Kinetics approach to subdiffusion essentially along the lines suggested by the CTRW model of a generalized walker and detail the various approximations. In Section 3 we present the potentialities of the Monte Carlo method in simulating subdiffusive walkers in various environments and conditions and describe the power law pdf from which we sample the rest times. In Section 4 we report the main results of the Monte Carlo simulations of a purely subdiffusive process and compare these results to the analytical ones deriving from the Fractional Kinetics approach. In Section 5 we show one of the possible extensions of the Monte Carlo approach: the simulation of an advection-dispersion process in a heterogeneous medium composed of two zones with different physical properties. In Section 6 we finally draw some conclusion of this work. Several appendices make this paper self-consistent.

## 2. Brief review of Fractional Kinetics

Let us consider the continuous time random walk (CTRW) of a particle in a medium, consisting of a sequence of jump-trapping-release processes: more specifically, the transport will be described in terms of the random walk of a walker which performs a succession of instantaneous jumps of length $x$ (drawn from a probability density function (pdf) $\lambda(x)$), followed by time intervals $t$ (drawn from a pdf $w(t)$) during which it is trapped at the location reached in the last jump. The assumption that these two probabilities do not depend on the past fate of the walker implies that in any case the process is semi-Markovian, i.e. Markovian at the jump instants. Better specification depends on the choice of $w(t)$ and of $\lambda(t)$.

Physical examples of such motion may be found e.g. in the underground transport of carriers of toxic or radioactive contaminants, subject to immobilizations in and releases from the solid environment, caused by the subsurface water flow (Kosakowski 2004), or in the transport of charge carriers in amorphous semiconductors under a constant electric field (Scher, Montroll 1975). Clearly, assuming linearity – i.e. absence of interactions between the walkers –, this phenomenological picture lends itself to be Monte Carlo (MC) simulated once the pdf's $\lambda(x)$ and $w(t)$ and the medium parameters are assigned. However, before describing the steps of the MC simulation, we prefer to summarize the analytical approach (Metzler, Klafter 2004; Metzler, Klafter 2000; Montroll, Weiss 1965), which enlightens many features of the underlying transport process.

The process may be formalized by introducing firstly a bivariate probability $p(x,t)dxdt$ that the walker arrives at a location $x$ at time $t$ (actually at a location between $x$ and $x+dx$ at a time between $t$ and $t+dt$) and then another univariate probability $P(x,t)dx$ that the walker is at $x$ (more precisely, between $x$ and $x+dx$) at time $t$, independently of when it has arrived at $x$. The pdf $p(x,t)$ may be expressed by a Chapman-Kolmogorov type integral equation in terms of the probabilities that the walker *i)* arrived at any location $x' \in (-\infty, \infty)$ at any previous time $t' \leq t$ and *ii)* remained immobilized at $x'$ for a time interval $t - t'$ and, finally, *iii)* performed an instantaneous jump from $x'$ to $x$:

$$p(x,t) = \int_0^t dt' \int_{-\infty}^{\infty} dx' \, p(x',t') w(t-t') \lambda(x-x') + \delta(x)\delta(t) \qquad (1)$$

where the second term at *rhs* represents the initial condition of the walk, namely that the walker started at $x=0$ at $t=0$.

The other pdf $P(x,t)$ may now be expressed as the integral of the product of the probability that the walker arrived at $x$ at any time $t' \leq t$, namely $p(x,t')dxdt'$, times the probability $\phi(t-t')$ that it remained immobilized there at least for $t-t'$, viz.,

$$P(x,t) = \int_0^t dt' \, p(x,t') \phi(t-t') \qquad (2)$$

where

$$\phi(t) = 1 - \int_0^t dt' \, w(t') \qquad (3)$$

An explicit expression for $P(x,t)$ might be obtained by firstly performing the Laplace- ($t \to u$) and the Fourier- ($x \to k$) transforms of eq. (2), thus obtaining $\hat{\tilde{P}}(k,u)$, and then by inverse transforming. We shall see that the latter step, with the aid of some approximations, leads to the diffusion or to the fractional diffusion equation, according to the choice of the pdf's $w(t)$ and $\lambda(x)$.

In the Laplace-Fourier space the relation (2) becomes (cf. eq. (A3) of the Appendix A):

$$\hat{\bar{P}}(k,u) = \frac{1-\bar{w}(u)}{u}\frac{1}{1-\bar{w}(u)\hat{\lambda}(k)} \qquad (4)$$

Note that this expression is rigorous, in the sense that it contains the same information as eqs. (1-3). To proceed further, we must specify the time and jump pdf's $w$ and $\lambda$. In the following, we shall consider the two cases which originate the diffusive and the subdiffusive transport, respectively.

Gaussian jumps and exponential waiting times: the diffusive case.
It is assumed that the walker performs Gaussian jumps with mean value zero and variance $\Sigma^2=2\sigma^2$. After each jump, it remains immobilized for an exponential time interval with mean value $\tau$.
The Gaussian pdf, its mean, variance and F-transform are

$$\lambda(x) = \frac{1}{\sqrt{2\pi}\Sigma}e^{-\frac{1}{2}\frac{x^2}{\Sigma^2}} \quad ; \quad E[x]=0 \quad ; \quad Var[x]=\Sigma^2 \equiv 2\sigma^2 \quad ; \quad \hat{\lambda}(k)=e^{-k^2\sigma^2} \qquad (5)$$

The exponential pdf, its mean, variance and L-transform are

$$w(t) = \frac{1}{\tau}e^{-\frac{t}{\tau}} \quad ; \quad E[t]=\tau \quad ; \quad Var[t]=\tau^2 \quad ; \quad \bar{w}(u)=\frac{1}{1+u\tau} \qquad (6)$$

Moreover, let $W(t)$ be the cumulative distribution function (cdf) of $w(t)$, viz. $W(t) = 1-e^{-\frac{t}{\tau}}$. Thus, the conditional probability $1-\frac{w(t)dt}{1-W(t)}$ for a rest time $t+dt$, given that a rest time $t$ has already elapsed, is $1-\frac{dt}{\tau}$. Then, the probability that the rest time ends at $t$, so that a step of the random walk will indeed take place in $(t, t+dt)$, as $dt \to 0$, is $1-\left(1-\frac{dt}{\tau}\right) = \frac{dt}{\tau}$, independently of $t$. Then, the assumed semi-Markovian process is actually Markovian.

Since both distributions have finite first and second moments, the central limit theorem applies (Metzler, Klafter 2000; Bouchaud, Georges 1990) and we know in advance that the probability $P(x,t)$ that the walker be at $x$ at time $t$ is Gaussian. Substituting the above transforms in eq. (4) yields

$$\tau(u\hat{\bar{P}}(k,u)-1) + (1-e^{-k^2\sigma^2})\hat{\bar{P}}(k,u) = 0 \qquad (7)$$

To the lowest order approximation in $k$, i.e. for large $x$, we approximate

$$\hat{\lambda}(k) = e^{-k^2\sigma^2} \approx 1-k^2\sigma^2 \qquad (8)$$

and then $\tau(u\hat{\bar{P}}(k,u)-1) + k^2\sigma^2\hat{\bar{P}}(k,u) = 0$. Inverse L-transforming and then inverse F-transforming yields the standard diffusion equation, viz.,

$$\frac{\partial P(x,t)}{\partial t} = D\frac{\partial^2 P(x,t)}{\partial x^2} \qquad (9)$$

where $D = \frac{\sigma^2}{\tau} = \frac{\Sigma^2}{2\tau}$ is the well known diffusion coefficient.

The variance of $P(x,t)$ is readily obtained by applying the operator $\int_{-\infty}^{\infty} x^2 dx$ to the diffusion equation

$$\frac{d}{dt}\langle x^2(t)\rangle = \frac{\sigma^2}{\tau}\int_{-\infty}^{\infty} x^2 \frac{\partial^2 P(x,t)}{\partial x^2}dx$$

Integrating twice by parts the integral at *rhs* yields 2 (recollect that $\int_{-\infty}^{\infty} dx P(x,t) = 1$), and finally

$$\langle x^2(t)\rangle = \frac{2\sigma^2}{\tau}t = \frac{\Sigma^2}{\tau}t = 2Dt \qquad (10)$$

In words, the variance of the diffusion process increases linearly with time.

The solution to the diffusion equation (9) is

$$P(x,t) = \frac{1}{\sqrt{4\pi Dt}} e^{-\frac{(x-x')^2}{4Dt}} \tag{11}$$

It is important to note that eq. (10) is rigorous in spite of the fact that it has been obtained by utilizing the approximation (8) of the F-transform of $\lambda(x)$. Indeed, back transforming eq. (7), yields

$$\tau \frac{\partial P(x,t)}{\partial t} = -P(x,t) + \int_{-\infty}^{\infty} dx' \lambda(x-x') P(x',t)$$

Application of the operator $\int_{-\infty}^{\infty} x^2 dx$, after some algebra concerning the convolution integral, yields

$\tau \frac{d}{dt} \langle x^2 \rangle = -\langle x^2 \rangle + \langle x^2 \rangle + \Sigma^2$ and then eq. (10).

Gaussian jumps and power law with α<1 waiting times: the subdiffusive case.
In the subdiffusive motion of a walker, e.g. in a porous medium (Montroll, Weiss 1965; Margolin, Berkowitz 2004), the sequence of flights and trappings is characterized by long rests, much longer than those occurring in the diffusive case. Formally, this means that the pdf $P(x,t)$ should have tails fatter than the Gaussian ones pertaining to the diffusive case and, correspondingly, that the pdf $w(t)$ (from which the waiting times are sampled) should be characterized by fat tails, fatter than the exponential ones of the diffusive case. Furthermore, it is required that the distribution of the sum of the waiting times – which represents the walker's temporal fate – should have tails not substantially variable during the whole walker motion, when new terms are continuously added. This implies that the mean of the sum of the waiting times, constituted by a sequence of several *iid* random variables sampled from $w(t)$, should have the same tails, up to a scale factor, as those of $w(t)$. In other words, this sum should give rise to a limiting stable distribution [1], similarly to the case of the distributions obeying the central-limit theorem, which roughly states that the limiting distribution of the sum of any *iid* random variables is Gaussian, provided that the first and second moments of the terms are finite. The upshot is that, to obtain a distribution which is stable and endowed with fatter than Gaussian tails, we should abandon the requirement of the moments finiteness. The solution to this problem is due to the Lévy and Khinchine generalization (Feller 1971) of the central-limit theorem, which takes into account the possibility that the terms of the sum obey distributions having infinite moments. This theorem specifies that a power law of the kind $w(t) \propto (\tau^\alpha / t^{1+\alpha})$, with $0 < \alpha < 1$, $\tau$ positive, small and $t \geq \tau$ to prevent divergence, characterized by infinite first moment, satisfies both the above requirements, namely that the mean of the sum of *n* terms thereby sampled will have fat tails and limiting stable distribution.

It is of paramount importance observing that it turns out that the subdiffusive experiments mentioned in the Introduction are actually well accounted for by power laws with suitable *α<1* exponents: thus, *mathematics matches physics*.
The Appendix B, after (Metzler, Klafter 2004; Metzler, Klafter 2000), gives some more details on the Lévy-Gnedenko generalized central-limit theorem.

In view of the following comparison between Monte Carlo simulations and analytical results, the formalization of the subdiffusive transport is performed by assuming the same Gaussian distribution (5) for the jump lengths and a pdf with power law asymptotic for the waiting times, viz.,

$$w_1(t) = \frac{2\alpha}{2+\alpha} \frac{1}{\tau^2} t = p \frac{2}{\tau^2} t = p\, w_1^*(t), \qquad t \in \Delta_1 t = (0, \tau)$$

$$w_2(t) = \frac{2\alpha}{2+\alpha} \tau^\alpha \frac{1}{t^{1+\alpha}} = (1-p)\tau^\alpha \frac{\alpha}{t^{1+\alpha}} = (1-p) w_2^*(t), \qquad t \in \Delta_2 t = (\tau, +\infty) \tag{12}$$

---

[1] (Feller 1971, p.166) Let $X, X_1, X_2, \ldots$ be a sequence of mutual independent *rv* obeying a common distribution *R* and let $S_n = X_1 + X_2 + \ldots + X_n$. The distribution *R* is said to be stable if $S_n$ and $c_n X$ have same distribution, viz., $S_n \stackrel{d}{=} c_n X$. A theorem states that only the constants $c_n = n^{\frac{1}{\alpha}}$, with $0 < \alpha \leq 2$, are possible. The constant α is called the characteristic exponent of *R*.

where $p = \dfrac{\alpha}{2+\alpha}$ is the probability of a waiting time less or equal to $\tau$ and the $w_i^*(t)$, $i=1,2$, are the pdf's normalized on the domains $\Delta_i t$. Moreover, let $W_i^*(t)$ be the cdf of $w_i^*(t)$, viz.

$$W_1^*(t) = \int_0^t w_1^*(v)\,dv = \frac{t^2}{\tau^2} \quad \text{and} \quad W_2^*(t) = \int_\tau^t w_2^*(v)\,dv = 1 - \frac{\tau^\alpha}{t^\alpha}.$$

In this case, the conditional probability $1 - \dfrac{w_i^*(t)\,dt}{1 - W_i^*(t)}$ for a rest time $t+dt$, given that a rest time $t$ has already elapsed, is $1 - \dfrac{2t\,dt}{\tau^2 - t^2}$ for $t \in \Delta_1 t$ and $1 - \alpha\dfrac{dt}{t}$ for $t \in \Delta_2 t$. Then, the probabilities that a step of the random walk will indeed take place in $(t, t+dt)$, as $dt \to 0$, are $\dfrac{2t\,dt}{\tau^2 - t^2}$ and $\alpha\dfrac{dt}{t}$, respectively, both depending on $t$. Then, the underlying process remains semi-Markovian, i.e. Markovian only at the jump instants. The initial linear shape of $w(t)$ has been here introduced to avoid the sudden jump from zero to $\alpha/\tau$, which would have occurred in case of a pure power law $w_2^*(t) = (\alpha/\tau)(\tau/t)^{1+\alpha}$ for $t \in (\tau, \infty)$. Correspondingly, due to the smallness of $\tau$, successive jumps essentially with no rest time are possible. The mean number of such jumps is $\dfrac{1}{1-p} = 1 + \dfrac{\alpha}{2}$.

The L-transform of the above pdf (12) is $\overline{w}(u) = \overline{w}_1(u) + \overline{w}_2(u)$, where

$$\overline{w}_1(u) = \frac{2\alpha}{2+\alpha}\frac{1-(1+u\tau)e^{-u\tau}}{(u\tau)^2} \quad \text{and} \quad \overline{w}_2(u) = \frac{2}{(2+\alpha)(1-\alpha)}\left[(1-\alpha+u\tau)e^{-u\tau} - (u\tau)^\alpha \int_{u\tau}^\infty x^{1-\alpha} e^{-x}\,dx\right]$$

To the first order in $u$, and then for large $t$, we approximate

$$e^{-ut} \cong 1 - ut \quad \text{and} \quad \int_{u\tau}^\infty x^{1-\alpha} e^{-x}\,dx \cong \int_0^\infty x^{1-\alpha} e^{-x}\,dx = \Gamma(2-\alpha)$$

so that

$$\overline{w}(u) = 1 + c_1 u\tau + c_\alpha (u\tau)^\alpha \tag{13}$$

where

$$c_1 = \frac{2\alpha}{(1-\alpha)(2+\alpha)} \quad \text{and} \quad c_\alpha = \frac{2\Gamma(1-\alpha)}{2+\alpha} \tag{14}$$

where $\Gamma(.)$ is the complete Gamma function.

Substituting the approximate eqs. (8) and (13) in the rigorous expression (4) we get an approximate expression for the Laplace - Fourier transform of the pdf $P(x,t)$ that the walker be in $x$ at time $t$

$$\hat{\overline{P}}(k,u) = \frac{\left[-c_1(u\tau)^{1-\alpha} + c_\alpha\right]\dfrac{1}{u}}{\sigma^2 k^2 (u\tau)^{-\alpha} - c_1(u\tau)^{1-\alpha} + c_\alpha} \tag{15}$$

Inverse L- and F- transforming, as detailed in the Appendix C, yields the required Fractional Diffusion Equation (FDE)

$$c_\alpha \frac{\partial}{\partial t} P(x,t) - \sigma^2 \tau^{-\alpha} {}_0D_t^{1-\alpha} \frac{\partial^2}{\partial x^2} P(x,t) - c_1 \tau^{1-\alpha} \frac{\partial}{\partial t} {}_0D_t^{1-\alpha} P(x,t) = -c_1 \frac{1}{\Gamma(\alpha-1)}\frac{1}{t}\left(\frac{\tau}{t}\right)^{1-\alpha} \delta(x) \tag{16}$$

Note that the presence of the integral operator ${}_0D_t^{1-\alpha}$ accounts for the non-locality of the above equation, which reflects the semi-Markovian feature of the process. In lack of an explicit solution for (16), eq. (15) is usually approximated (Metzler, Klafter 2000; Metzler, Klafter 2004) by expanding to the lowest order $u^\alpha$ the L-transform of the waiting time $w(t)$. Then, $c_1 = 0$. As pointed out in (Margolin, Berkowitz 2004) and as we shall see in the next Section devoted to the use of the Monte Carlo simulation for validating the approximate analytical approach, the term linear in $u$ becomes important at long times only when α is close to *1*. By neglecting the linear term, eq. (16) has an explicit solution in the F- and L- transformed space, namely the Fox function $P(x,t)$, which can be expressed in a computable form (Metzler, Klafter 2000; Metzler, Klafter 2004; Margolin, Berkowitz 2004):

$$P(x,t) = \frac{1}{\sqrt{4K_\alpha t^\alpha}} \sum_{n=0}^{+\infty} \frac{(-1)^n}{n!\Gamma\left(1-\frac{\alpha(n+1)}{2}\right)} \left(\frac{x^2}{K_\alpha t^\alpha}\right)^{\frac{n}{2}} \quad (17)$$

As in the case of the standard diffusion, the most important parameter characterizing the transport is the variance $\langle x^2(t)\rangle$. A fractional differential equation in time for this quantity is readily obtained by applying the operator $\int_{-\infty}^{\infty} x^2 dx$ to eq. (16). In the analytical derivation, the most interesting algebraic step is that by integrating twice by parts $\int_{-\infty}^{\infty} x^2 \frac{\partial^2}{\partial x^2} P(x,t)dx$ simply yields $2$, so that, by eq. (C4') of the Appendix C, $\left[_0 D_t^{1-\alpha} 2\right] = 2t^{\alpha-1}/\Gamma(\alpha)$. Therefore,

$$\frac{d}{dt}\langle x^2(t)\rangle = \frac{2K_\alpha}{c_\alpha}\frac{t^{\alpha-1}}{\Gamma(\alpha)} + \frac{c_1}{c_\alpha}\tau^{1-\alpha}\frac{d}{dt}{}_0 D_t^{1-\alpha}\langle x^2(t)\rangle \quad (18)$$

where $K_\alpha = \sigma^2/\tau^\alpha = \Sigma^2/2\tau^\alpha$ is the generalized diffusion coefficient.

Then, by expanding to the lowest order $u^\alpha$ the L-transform of the waiting time $w(t)$ as done before, in eq. (18) $c_1 = 0$ and the second term at *rhs* disappears. Neglecting this term, the above differential equation becomes no more fractional and its solution is

$$\langle x^2(t)\rangle = \frac{1}{c_\alpha}\frac{2K_\alpha}{\Gamma(1+\alpha)}t^\alpha \quad (19)$$

which differs from those reported in literature, e.g. (Metzler, Klafter 2000; Metzler, Klafter 2004), for the presence of the coefficient $c_\alpha$, which grows from the initial value *1* for *α=0*, to 2 for *α=2/3* and to 6.6 for *α=0.9*. The above expression (19) for the mean squared displacement of the walker at the generic time *t* confirms the subdiffusive character of the transport when $0 < \alpha < 1$ - i.e. the spatial spread of the walker around its initial position grows sublinearly with time - and represents one of the most important findings of the fractional diffusion approach. To evaluate the relevance of the approximation leading to (19), we come back to equation (18), which may be analytically transformed in the Laplace space, leading to the simple algebraic expression

$$\langle x^2(u)\rangle = \frac{2\sigma^2}{c_\alpha u^{\alpha+1}\tau^\alpha - c_1 u^2 \tau} \quad (20)$$

This expression has the important advantage of lending itself to numerical Laplace inversion. To do this, we have adopted the algorithm (Hollenbeck 1998). The curve thus obtained, which accounts also for the contribution of the linear term, will be used in Section 4 in comparison with the MC variance and the analytical approximated expression (19): we anticipate that (19) will turn out to be rather correct up to values $\alpha \approx .6$, while (20) holds true over the entire range of α. Indeed, it can be observed that in eq. (20), as $\alpha \to 1$, the divergence of the coefficient $c_\alpha$ is exactly compensated by the one of $c_1$: by adopting (19), this compensation is lost and the variance will significantly be underestimated for large values of *α*.

## 3. The Monte Carlo approach to the continuous time random walk (CTRW)

The MC approach to the particle transport is based on the simulation of the *microscopic* behaviour of an ensemble of particles, evolving in phase space according to a master equation such as eq. (1) and *macroscopically* resulting in the behaviour of the ensemble. In the case of the anomalous transport, a distinct merit of the MC approach with respect to the analytical one of the fractional kinetics is that the simulation is based on the exact $w(t)$ and $\lambda(t)$ pdf's governing the fate of each particle, without the necessity of resorting to the various approximations of these pdf's as detailed in the preceding Section. Therefore, the MC simulations may be utilized to quantitatively evaluate the accuracy of the analytical results. Moreover, the MC approach lends itself to examine a number of variants of the walk not easily affordable analytically. For example, it is not necessary that the pdf's of the waiting times and of the jump lengths be independent each from the other and therefore also correlated walks may be easily studied. Indeed, if $\psi(x,t)$ is the bivariate pdf of the jump lengths and trapping times, we sample a jump length $X$ from the marginal $\lambda(x) = \int_0^\infty \psi(x,t)dt$ and then a waiting time from the conditional $w(t) = \psi(t|X) = \dfrac{\psi(t,X)}{\lambda(X)}$.

In addition, the medium in which the walker is moving may be as well anisotropic, with pdf's containing space and time dependent parameters, e.g. as in Section 5. Finally, the transport is easily simulated in 2D and even 3D spaces. As in almost all cases, the main drawbacks of the MC approach are the requirements of long CPU times and vast memory occupation: however, as time goes by, the continuous computer improvements make these burdens less and less important.

In a computer code, before starting the stochastic simulation, the extension of the phase space available for the walker, namely the available space $(x_{min}, x_{max})$ and time $t_{max}$, must be fixed. Then, this rectangular domain is discretized by assigning the respective numbers of subintervals $N_x$ and $N_t$: correspondingly a matrix $P$ of order $(N_x, N_t)$ is assigned, whose generic element $P(i,j)$ at the end of the simulation will contain the number of passages of the walker in the elementary $ij$-th cell. To avoid distortions in the bivariate distribution $P$ and then gross errors in the mean and variance as a function of time, it is important that the space assignment must be done so that essentially no walks will exit from the space interval before $t_{max}$.

The simulation of the fate of each particle is performed by sampling the successive jump lengths $\Delta x$ and waiting times $\Delta t$ from the pdf's (5) and (12), respectively. Specifically, $\Delta x = \Sigma r_G$, where $r_G$ is a normal random variable (rv) with zero mean and unit variance, i.e. $r_G \sim N(0,1)$. Concerning $\Delta t$, the cdf of the $w_i(\Delta t)$ given by eq.(12) are

$$W_1(\Delta t) = p W_1^*(t) = \dfrac{\alpha}{2+\alpha}\left(\dfrac{\Delta t}{\tau}\right)^2 \quad \text{for} \quad \Delta t \leq \tau$$

$$W_2(\Delta t) = p W_1^*(\tau) + (1-p)W_2^*(\Delta t) = 1 - \dfrac{2}{2+\alpha}\left(\dfrac{\tau}{\Delta t}\right)^\alpha \quad \text{for} \quad \Delta t \geq \tau \qquad (21)$$

To sample a $\Delta t$ value, a rv $r_U$ is firstly sampled from the uniform distribution, i.e. $r_U \sim U[0,1)$ and then this value is equated to $W(\Delta t)$. Solving with respect to $\Delta t$ yields

$$\Delta t = \tau\sqrt{\dfrac{r_U(2+\alpha)}{\alpha}} \quad \text{if} \quad r_U \leq \dfrac{\alpha}{2+\alpha} \quad ; \quad \Delta t = \tau\left[\left(\dfrac{2+\alpha}{2}\right)(1-r_U)\right]^{-\frac{1}{\alpha}} \quad \text{if} \quad r_U > \dfrac{\alpha}{2+\alpha} \qquad (22)$$

In *Figures 1* and *2* we show the characteristic shapes of MC simulated walks in the *x-t* space, with different values of α: note that for *α=.3* the motion is strongly subdiffusive and indeed the story is characterized by long rest times. The other story, with *α=.99*, is instead very close to the one of a fully diffusive motion *(α=1)*.

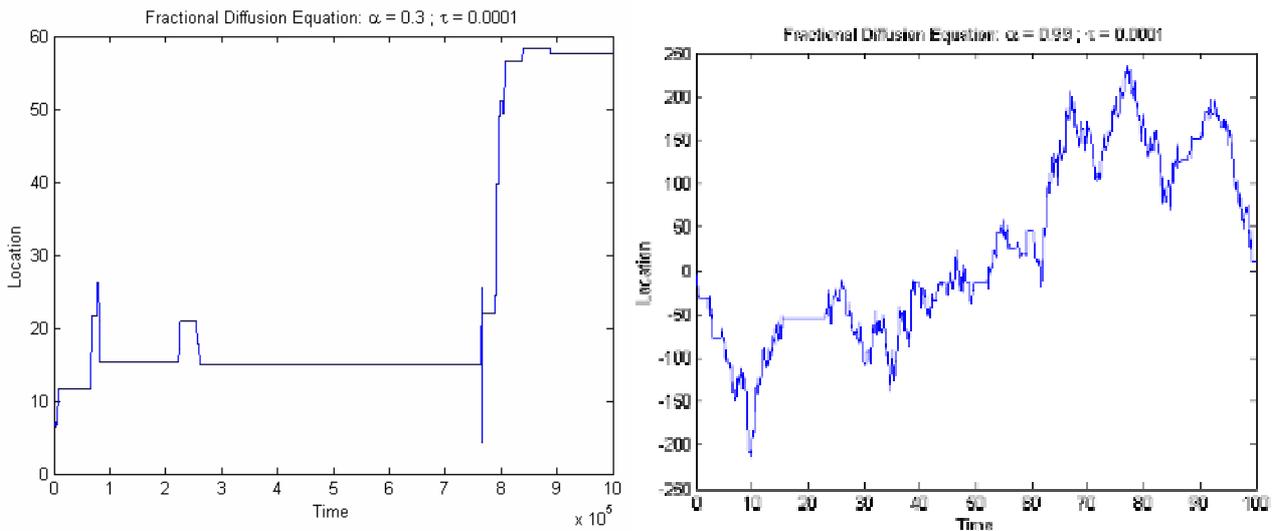

*Figures 1 and 2. Characteristic shapes of walks in phase space x-t, for different values of α. Left: α=.3; right: α=.99.*

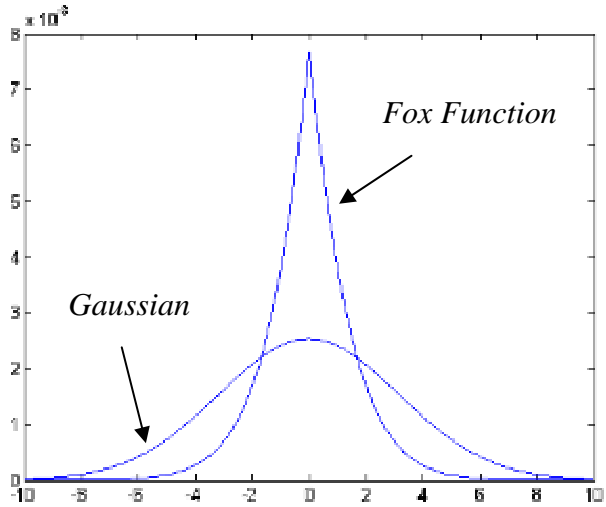

*Figure 3. Typical shapes of a Fox function (tent-like, $\alpha=.5$) and a Gaussian, at the same time.*

At the generic step of the simulation, the walker arrives at a point *(x,t)* within the cell *(i,j)* and the new $\Delta x$ and $\Delta t$ are successively sampled. This implies that the walker remains in the *i*-th spatial interval until $t'=t+\Delta t$ and then it suddenly performs a spatial jump towards the new point $(x' = x + \Delta x, t')$ within the cell $(i + n_i, j + m_j)$ with $n_i$ and $m_j$ integers. Then, a unit is added to all the elements $(i, j), (i+1, j),...(i+(n_i -1), j)$ of the matrix *P* and the walk continues from *(x',t')*. When a large number of stories have been processed in this way and the walker passages through the discrete cells have been registered, the normalized space-time matrix *P(i,j)* will represent the (discretized) distribution of the particles: in particular, the *j-th* column of the matrix represents the spatial distribution at time *j*. In *Figure 3* we compare the results for *P(x,t)* at a given time in the case of a Fox function (cf. eq. (17)) with $\alpha=.5$ and a Gaussian function ($\alpha=1$): observe that the Fox function shows greater persistence of the initial condition and a smaller spread around the origin with respect to the Gaussian.

**4. Results and comparisons for a subdiffusive walker**

In this Section we present a comparison between the results of the analytical (and numerical) FDE approach to subdiffusion and those of the Monte Carlo simulations both for the spatial distribution $P(x,t)$ at given times *t* and for the time evolution of the variance for different values of the parameter $\alpha$.

<u>Case 1</u>. The parameters values are: $\alpha=0.3$, $\tau=10^{-4}$, $\sigma=.5$. The number of simulated stories is $10^5$ and each story will end if the maximum time $t_{max}=10^4$ is reached or if the boundaries of the spatial domain are attained during the simulation. These boundaries are allocated through an *a priori* estimate of the mean spread of the location *x* given by several $\sqrt{\langle x^2(t_{max})\rangle}$.

*Figure 4* reports the spatial distribution *P(x,t)* at a given time as obtained from the MC simulation and from the analytical Fox function in its representation by series (17). *Figure 5* reports the variances as a function of time, as obtained from the analytical expression (19), from the numerically inverted curve (20) and from the MC. Since in this case $\alpha$ is small, we expect that the effects of the linear term neglected in (19) should be almost negligible and indeed in all cases the analytical curves are in very good agreement with the reference ones of the MC simulation.

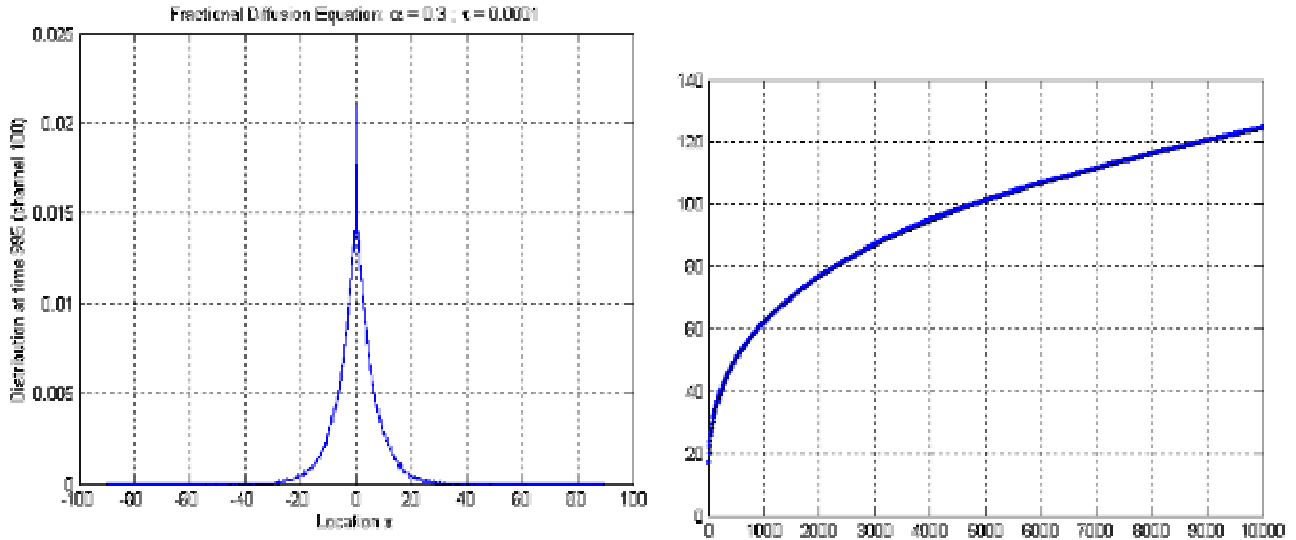

*Figures 4 and 5. Spatial distribution P(x,t) for a given time(left) and variance evolution in time (right).*

Case 2. The parameters values are: $\alpha=0.5$, $\tau=10^{-2}$, $\sigma=.5$. The number of simulated stories is $10^5$ and each story will end if the maximum time $t_{max}=10^4$ is reached or if the boundaries of the spatial domain are attained during the simulation. In *Figures 6* and *7* we show the spatial distribution at a given time and the variance, as above. We remark that also in this case, while $\alpha$ has been increased to .5, the contribution of the linear term is negligible in the time range considered, so that the analytical results still coincide with the reference MC values.

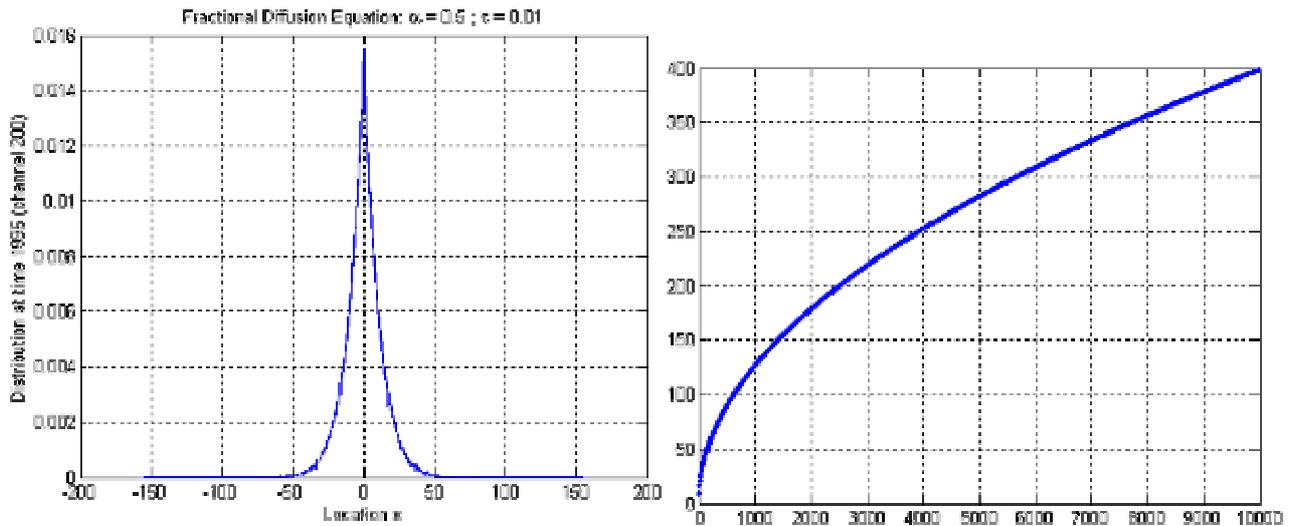

*Figures 6 and 7. Spatial distribution P(x,t) for a given time(left) and variance evolution in time (right).*

Case 3. The parameters values are: $\alpha=0.8$, $\tau=1$, $\sigma=.5$. The number of simulated stories is $10^5$ and each story will end if the maximum time $t_{max}=10^4$ is reached or if the boundaries of the spatial domain are attained during the simulation. *Figure 8* shows the analytical and the MC spatial distributions at a given time. *Figure 9* displays an enlarged representation of one of the (symmetric) tails of the *P(x,t)* shown in *Figure 8*. It appears that the analytical distribution slightly but systematically underestimates the reference MC profile. In *Figures 10* and *11* the analytical and MC variances are plotted in log-log and linear scale as a function of time. We remark that in this case ($\alpha$ close to *1*) the contribution of the effects of the linear term are no more negligible. However, when the variance given by expression (19) is replaced by the numerical inversion of expression (20), which contains also the linear term, the data perfectly agree with the MC ones, except at small times (cf. *Figures 10* and *12*). We conjecture that these small discrepancies could be due to the so called "*diffusion limit*" approximations (i.e. the expansion of the exact expressions of $\hat{\lambda}(k)$ and $\overline{w}(u)$ for small *k* and small *u*, respectively), not appearing in the reference MC approach. We note also that the discrepancy between the FDE analytical model and the reference MC (which does not suffer from approximations) is by far more evident in the variance than in the spatial distribution plots. This happens because of the large contributions of the tails of *P(x,t)* to the variance.

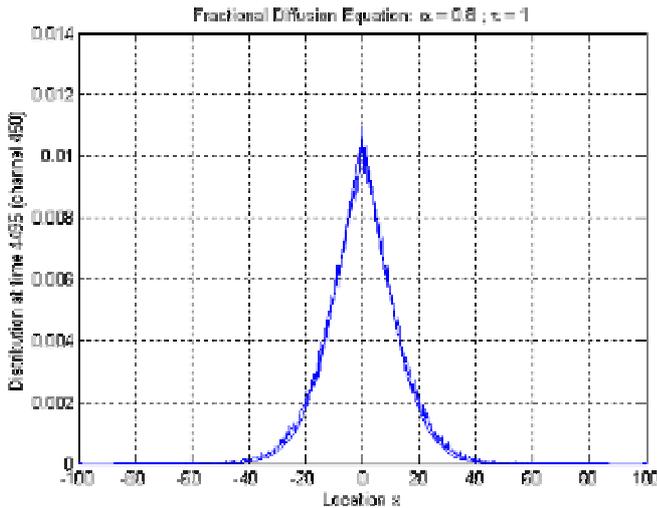 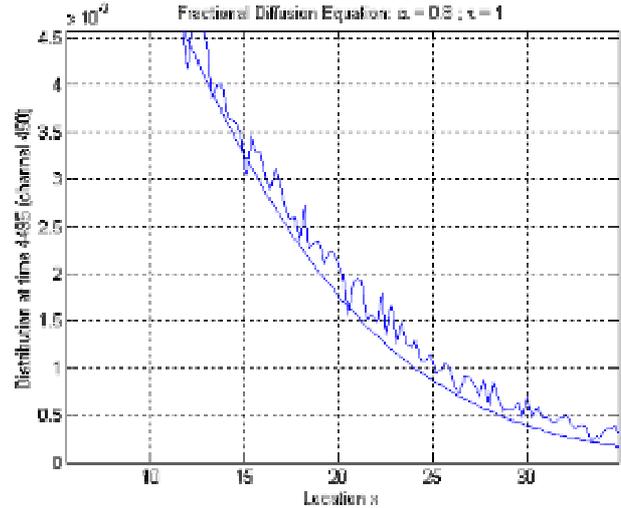

*Figures 8 and 9. Spatial distribution P(x,t) for a given time(left) and zoom of its right tail (right).*

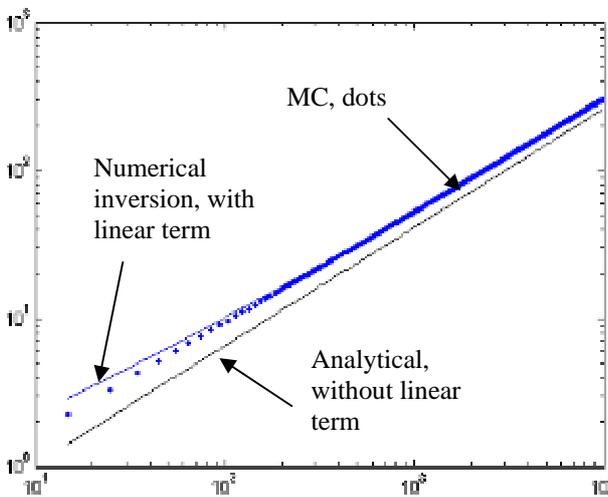 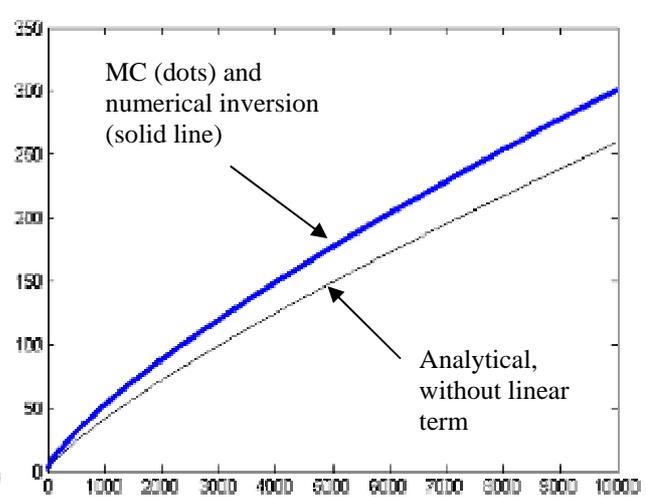

*Figures 10 and 11. Double logarithmic (left) and linear (right) scale plot of the variance vs. time.*

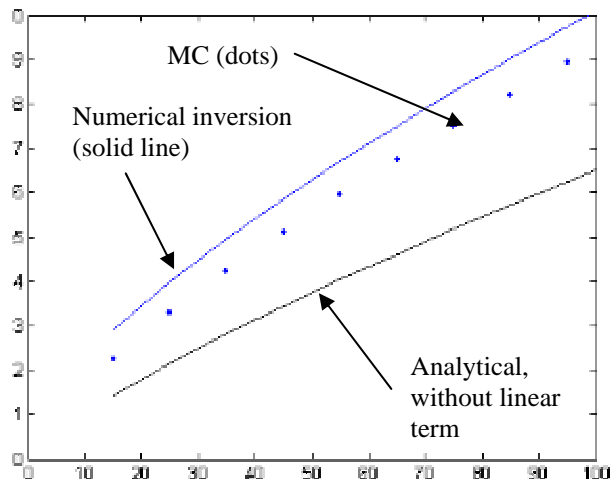

*Figure 12. Zoom of Figure 11 left, to enlighten the (small) discrepancy between MC and numerically inverted curves*

**5. Generalization of the anomalous advection-dispersion model for a two-layers medium**

We now generalize the stochastic model so far described to the case of the anomalous transport in heterogeneous media. This problem is of importance in different instances, since in reality the transport of particles, e.g. under the influence of an external constant force, gravitational or electric, or the underground transport of radioactive or toxic particles driven by the subsurface water flow normally occurs in heterogeneous media. In the simpler case of an homogeneous medium the problem is analytically tackled by means of the fractional Fokker-Planck equation or by means of the formally equal fractional advection equation (van Kampen 1981). Here we focus on the underground transport of particles.

We assume that the diffusion of the walker occurs within a constant velocity field v, in a medium constituted by two different homogeneous layers, say $L_1$ on the left and $L_2$ on the right, the discontinuity occurring at $x = x_d$. In layer $i$ ($i$=1,2) the diffusion contribution to each jump is Gaussian $(0, \Sigma_i^2)$ and the time intervals $\Delta t$ between jumps obey to a power law distribution $w_i$ with parameter $\alpha_i$ (cf. eq. (12)). In both layers the advection contribution to the walker motion is $v\Delta t$. Differently from the sudden-jump model adopted in the preceding Sections, we now assume that during the time between successive jumps the walker travels with constant speed, variable from jump to jump, given by the ratio between the total jump length and the waiting time. For sake of simplicity, here we assume also the same parameter $\alpha$ for the two zones. Moreover, to a first approximation, in the MC simulation it may be assumed that the optical path related to each Gaussian jump across the discontinuity can be subdivided proportionally to the respective standard deviations.

Let us now consider a single jump starting at $x_0$ in $L_1$. The contribution to the walker's travel due to diffusion is the so-called optical path $r_G \Sigma_1$, $r_G \sim N(0,1)$, and the jump duration is $T$, sampled from the power law $w_1$. If the whole jump is done within the same $L_1$, i.e. if $r_G \Sigma_1 + vT < x_d - x_0$, the walker location at the generic $t \leq T$ is (note that $r_G$ may well be negative)

$$x(t) = x_0 + (r_G \Sigma_1 + vT)\frac{t}{T} \qquad (23)$$

Otherwise, if $r_G \Sigma_1 + vT > x_d - x_0$, the walker crosses $x_d$ at time

$$t^* = \frac{x_d - x_0}{r_G \Sigma_1 + vT}T$$

and the utilized fraction of $r_G$ is

$$r_G \frac{t^*}{T} = \frac{x_d - x_0 - vt^*}{\Sigma_1}$$

At $t^*$ the walker is at the entrance of $L_2$. The remaining portion of the path within this layer is obtained from the residual fraction of $r_G$, viz.,

$$r_G \frac{t - t^*}{T} = \frac{x(t) - x_d - v(t - t^*)}{\Sigma_2}$$

so that

$$x(t) = x_d + (\Sigma_2 r_G + vT)\frac{t - t^*}{T} \qquad t^* \leq t \leq T$$

If $r_G$ is so negative that $\Sigma_2 r_G + vT < 0$, then $x(t) < x_d$, which would mathematically mean that the walker, as soon as arrived at $x_d$ driven by $v$, instead of proceeding further, is driven back to layer 1. Here, the advection velocity pushes it again towards layer 2 and these back and forth oscillations continue up to time $T$ with the walker essentially at rest in $x_d$. To take into account this possibility, the above equation is modified as follows

$$x(t) = x_d + [\max(0, \Sigma_2 r_G + vT)]\frac{t - t^*}{T} \qquad t^* \leq t \leq T \qquad (24)$$

Let us consider the other case, with the starting point $x_0$ of the jump in $L_2$, i.e. $x_0 > x_d$. Now the situation is reversed: if $x_0 + r_G \Sigma_2 + vT > x_d$, the whole jump occurs in this layer and the walker location at $t \leq T$ is

$$x(t) = x_0 + (r_G \Sigma_2 + vT)\frac{t}{T} \qquad (23')$$

Otherwise, if $x_0 + r_G \Sigma_2 + vT < x_d$, (which means that $r_G$ is so negative that also $r_G\Sigma_2+vT<0$) the walker flies backwards against the velocity field and crosses $x_d$ at time

$$t^* = \frac{x_d - x_0}{r_G \Sigma_2 + vT}T \qquad \text{or} \qquad t^* = \frac{x_0 - x_d}{|r_G \Sigma_2 + vT|}T$$

The utilized fraction of $r_G$ is

$$r_G \frac{t^*}{T} = \frac{x_d - x_0 - vt^*}{\Sigma_2} \qquad \text{or} \qquad |r_G|\frac{t^*}{T} = \frac{x_0 - x_d + vt^*}{\Sigma_2}$$

The remaining of the path within $L_2$ is obtained as before from the residual fraction of $r_G$, viz.,

$$x(t) = x_d + \left(\Sigma_1 r_G + vT\right)\frac{t-t^*}{T}$$

However, it may happen that $\Sigma_1 r_G + vT > 0$ so that, as in the preceding case, the walker is stuck at $x_d$. Application of an analogous remedy finally yields

$$x(t) = x_d + \min\left(0, \Sigma_1 r_G + vT\right)\frac{t-t^*}{T} \qquad t^* \leq t \leq T \qquad (24')$$

The following *Figures 13* and *14* show the spatial distribution $P(x,t)$ at two given times as obtained through MC simulation. The walkers move across two distinct zones, separated by a discontinuity at $x_d = 100$. The parameters of the two regions are the following: $\alpha_1=\alpha_2=.5$, $\sigma_1=5$, $\sigma_2=15$, $\tau=.01$. The flow speed $v$ was set to *30*, towards the right. As expected, the most relevant consequence of the medium heterogeneity is the significant discontinuity of $P(x,t)$ at $x_d$, due to the ratio of the walkers' speeds during their crossings of the discontinuity.

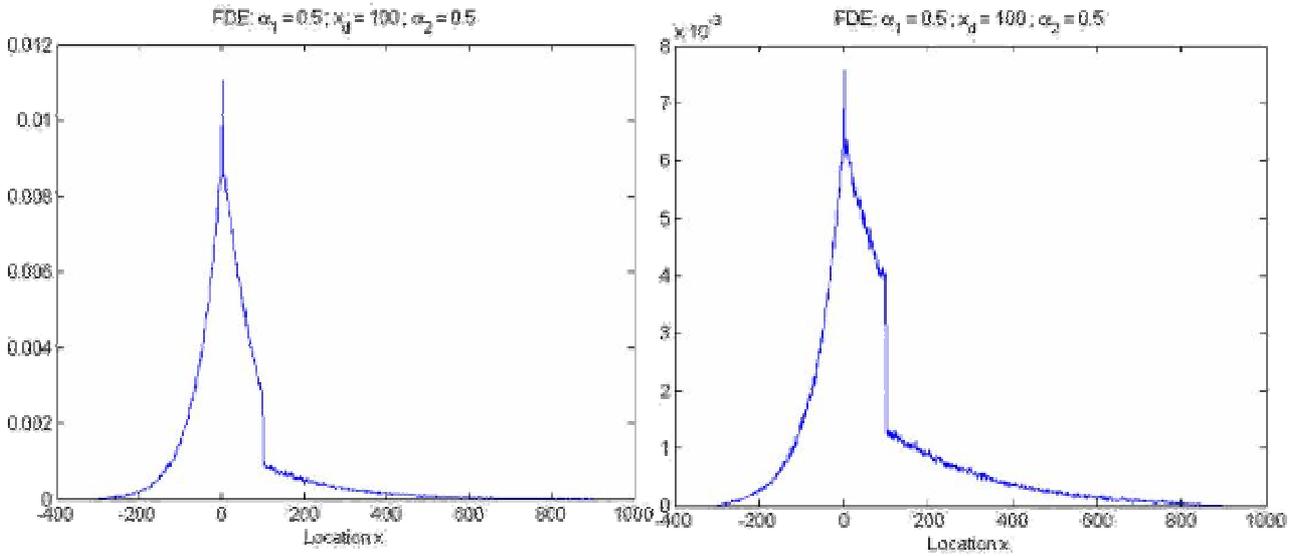

*Figures 13 and 14. Space distribution at time t=1 (left) and t=3 (right) for the model presented above.*

## 6. Conclusions

Several experimental investigations concerning transport of particles in given media have shown strong evidences for anomalous non-Fickian diffusion (Kosakowski 2004; Margolin, Berkowitz 2004; Metzler, Klafter 2000; Metzler, Klafter 2004; Scher, Montroll 1975). The interpretation of these results has given rise to various analytical models, which have finally led to the so-called Fractional Diffusion Equation (FDE) approach. Here, the motion of the particles in the surrounding medium is characterized by long sojourn times obeying distributions with infinite mean value, for which the standard central limit theorem must be replaced by the Lévy – Gnedenko generalised central limit theorem (Feller 1971; Metzler, Klafter, 2000). More specifically, the time fate of the particles is described in terms of a power law pdf with exponent $\alpha+1$: for the case here considered (i.e. $0<\alpha<1$), the mean squared displacement of the particles evolves as $t^\alpha$, thus resulting in a displacement smaller than the linear one characteristic of Fickian diffusion. Hence the name of subdiffusion given to such phenomena. The FDE analyses are performed in the Laplace and Fourier domains and a closed-form solution in the usual *x-t* space can not be obtained without resorting to several approximations. Moreover, the effects of the above mentioned approximations can hardly be physically interpreted in the reciprocal domain. Of utmost importance is that a term linear in *u* (where *u* is the L-transformed time variable) is not taken into account.

In this paper we propound the use of the Monte Carlo (MC) simulation, which does not suffer from any approximation (apart from those due to the finite statistics and to the *x-t* discretization), as a reference model in the investigation of the anomalous diffusion. The main result here attained is showing that the FDE analytical solutions are far from reality for $\alpha \to 1$. Nevertheless, the contribution of the above linear term may be taken into account by hybridizing the elegant fully analytical FDE approach with a numerical Laplace inversion. In this case, the resulting curves are quite consistent with the MC results. Another application, which follows from the MC flexibility, concerns the simulation of a more realistic anomalous advection-dispersion model in heterogeneous media, for which the analytical results are awkward to obtain. In lack of an analytical formulation, the MC approach could represent a viable way for numerical investigations of the particles behaviour at the interface of media characterized by different physical parameters. Further exploitations of the MC flexibility, not reported here, concern straightforward generalizations of the anomalous diffusion scheme, such as particles transport with pdfs correlated in time and space and transport in multidimensional media.

**Appendix A:** Laplace – Fourier transforms of equation (2)

The Laplace transform (L-transform) with respect to time $(t \to u)$ of eq. (1) is

$$\overline{p}(x,u) = \int_0^\infty dt\, e^{-ut} \int_0^t dt' \int_{-\infty}^\infty dx'\, p(x',t') w(t-t') \lambda(x-x') + \delta(x)$$

Inversion of the first and second integral yields

$$\overline{p}(x,u) = \int_0^\infty dt'\, e^{-ut'} \int_{t'}^\infty dt\, w(t-t') e^{-u(t-t')} \int_{-\infty}^\infty dx'\, p(x',t') \lambda(x-x') + \delta(x)$$

The inner integral is $\overline{w}(u)$, the L-transform of $w(t)$ and may be taken out of the triple integral. Inversion of the remaining two integrals yields

$$\overline{p}(x,u) = \overline{w}(u) \int_{-\infty}^\infty dx'\, \lambda(x-x') \int_0^\infty dt'\, p(x',t') e^{-ut'} + \delta(x)$$

The second integral is $p(x',u)$, the L-transform of $p(x',t)$, so that we get

$$\overline{p}(x,u) = \overline{w}(u) \int_{-\infty}^\infty dx'\, \lambda(x-x') \overline{p}(x',u) + \delta(x)$$

This expression is now Fourier transformed (F-transform) with respect to position $(x \to k)$, viz.,

$$\hat{\overline{p}}(k,u) = \overline{w}(u) \int_{-\infty}^\infty dx'\, \overline{p}(x',u) e^{-jkx'} \int_{-\infty}^\infty dx\, \lambda(x-x') e^{-jk(x-x')} + 1 = \overline{w}(u) \hat{\overline{p}}(k,u) \hat{\lambda}(k) + 1$$

Solving for $\hat{\overline{P}}(k,u)$ yields

$$\hat{\overline{p}}(k,u) = \frac{1}{1 - \overline{w}(u) \hat{\lambda}(k)} \tag{A1}$$

We now proceed to transform $P(x,t)$ as given by eqs. (2) and (3). The L-transform is

$$\overline{P}(x,u) = \int_0^\infty dt\, e^{-ut} \int_0^t dt'\, p(x,t') - \int_0^\infty dt\, e^{-ut} \int_0^t dt'\, p(x,t') \int_0^{t-t'} dt''\, w(t'') \to I - II \tag{A2}$$

Inversion of the integrals in the first term yields

$$I \to \int_0^\infty dt'\, p(x,t') \int_{t'}^\infty dt\, e^{-ut} = \frac{1}{u} \int_0^\infty dt'\, p(x,t') e^{-ut'} = \frac{\overline{p}(x,u)}{u}$$

Inverting the last two integrals in the second term of eq. (A2) yields

$$II \to \int_0^\infty dt\, e^{-ut} \int_0^t dt''\, w(t'') \int_0^{t-t''} dt'\, p(x,t')$$

Inverting the first and the second integrals in the above expression yields

$$II \to \int_0^\infty dt''\, w(t'') \int_{t''}^\infty dt\, e^{-ut} \int_0^{t-t''} dt'\, p(x,t')$$

Inverting the last two integrals yields

$$II \to \int_0^\infty dt''\, w(t'') \int_0^\infty dt'\, p(x,t') \int_{t'+t''}^\infty dt\, e^{-ut} = \frac{1}{u} \int_0^\infty dt''\, w(t'') \int_0^\infty dt'\, w(x,t') e^{-u(t'+t'')}$$

$$= \frac{1}{u} \int_0^\infty dt''\, w(t'') e^{-ut''} \int_0^\infty dt'\, p(x,t') e^{-ut'} = \frac{\overline{w}(u) \overline{p}(x,u)}{u}$$

Substitution of *I* and *II* in eq. (A2) yields

$$\overline{P}(x,u) = \frac{1-\overline{w}(u)}{u}\overline{p}(x,u)$$

Finally, F-transforming, we obtain the required LF-transform of $P(x,t)$, viz.,

$$\hat{\overline{P}}(k,u) = \frac{1-\overline{w}(u)}{u}\hat{\overline{p}}(k,u) = \frac{1-\overline{w}(u)}{u}\frac{1}{1-\overline{w}(u)\hat{\lambda}(k)} \quad (A3)$$

**Appendix B:** the Lévy – Gnedenko generalization of the central-limit theorem

Let $G(x;\mu,\sigma^2)$ be the normal (Gauss) pdf with mean value μ and variance $\sigma^2$. We say that the random variable (rv) ξ is normal (μ,σ) if

$$\Pr\{x \leq \xi \leq x+dx\} = G(x;\mu,\sigma^2)dx$$

If $n$ values $\xi_1, \xi_2, ..., \xi_n$ are independently sampled from the same $G$, the simplest version of the central limit theorem states that the rv $S_n = \frac{1}{\sqrt{n}}\sum_{i=1}^{n}(\xi_i - \mu) + \mu$ obeys to the same G, that is

$$\Pr\{x \leq S_n \leq x+dx\} = G(x;\mu,\sigma^2)dx$$

To simplify the matter, from now on we assume $\mu = 0$ and $\sigma = 1$. Then the theorem states that, up to a scale factor $n^{\frac{1}{2}}$, the sum of $n$ normal rv (0,1), namely $\sum_{i=1}^{n}\xi_i$ obeys to the same distribution as the one of the single terms of the sum. Formally, the distribution of the *rv* ξ is a stable distribution (see *Note 1*, § *2*) with characteristic exponent α=2. Actually, the theorem is much more general but here we only mention the above simple version.

When dealing with a power law pdf such as the *w(t)* given by eqs. (12), the situation is complicated by the lack of a finite mean value. In this case we may resort to the Gnedenko – Lévy generalization (Metzler, Klafter 2000) of the central limit theorem, which states that if the sum $S_n$ of $n$ iid (independent-identically-distributed) random variables $\xi_1, \xi_2, ..., \xi_n$, apart from an appropriate normalization, converges to some distribution *w(t)* in the limit $n \to \infty$, then *w(t)* is stable. Then, from the definition of stability

$$S_n \equiv \sum_{i=1}^{n}\xi_i \stackrel{d}{=} n^{\frac{1}{\alpha}}\xi$$

in the limit of large *n*. The following *Figures*, obtained from a MC simulation with $N = 10^6$ stories for the cases of *n=10* and *α=0.2* (*Figures 15* and *16*, case *a*) and *α=0.8* (*Figures 17* and *18*, case *b*) give an idea of how much relatively small *n* and *t* values are suitable to satisfy the theorem.

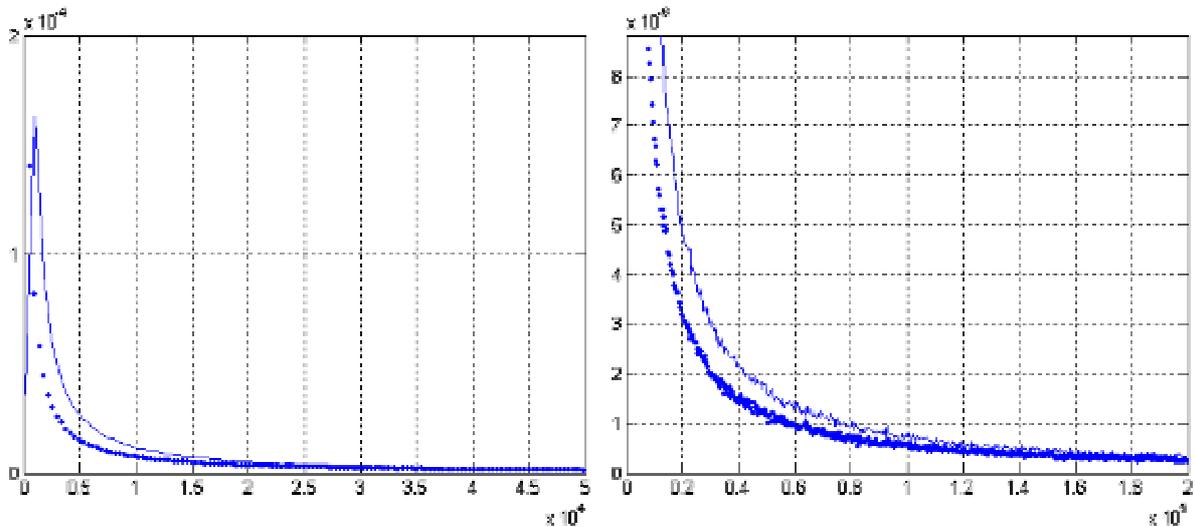

*Figures 15 and 16. a). Monte Carlo approximation of the Levy – Gnedenko theorem for finite n and t values. Dots= $S_n$; solid line= $n^{\frac{1}{\alpha}}\xi$. Left: plot of the distributions. Right: zoom. In this case α=.2*

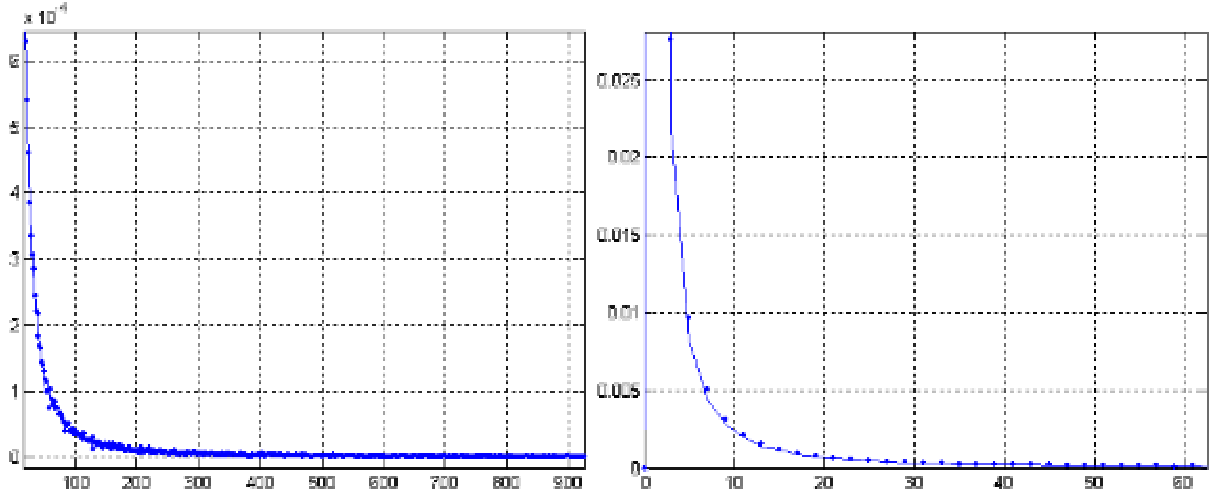

*Figures 17 and 18. b). The same as above. Left: plot of the distributions. Right: zoom. In this case α=.8*
*Note that the convergence is faster in this latter case, where α is closer to 1.*

**Appendix C:** The fractional diffusion equation (FDE)

For convenience, we firstly briefly recollect some definitions and properties of the fractional differ-integration. Detailed books on this subject are e.g. (Oldham, Spanier 1974; Miller, Ross 1993).

i) *Fractional integral*:
The Riemann-Liouville fractional integral of arbitrary order α, i.e. the integral operator $_0D_t^{-\alpha}$, is defined as

$$_0D_t^{-\alpha} f(t) = \frac{1}{\Gamma(\alpha)} \int_0^t (t-t')^{\alpha-1} f(t') dt' \tag{C1}$$

Note that the application of the operator to a function yields a result which has a power law memory of the previous function values. An important property of the fractional integral which follows from the fact that it is a convolution, is that its L-transform is the product of the transforms of $t^{\alpha-1}$ and $f(t)$, viz.,

$$L[_0D_t^{-\alpha} f(t)] = L[t^{\alpha-1}] \bar{f}(u) = u^{-\alpha} \bar{f}(u) \tag{C2}$$

ii) *Fractional derivative*:

Let $n = [\alpha]+1$, where $[\alpha]$ is the integer part of α. Differentiating (C1) $\nu \leq n$ times yields $\frac{d^\nu}{dt^\nu} {}_0D_t^{-\alpha} = {}_0D_t^{\nu-\alpha}$. If we change notations by performing the substitution $\nu - \alpha \to \alpha$ so that $0 \leq \alpha < 1$, we get the fractional derivative

$$_0D_t^\alpha = \frac{d^\nu}{dt^\nu} {}_0D_t^{\alpha-\nu} \quad \text{or, explicitly,} \quad {}_0D_t^\alpha f(t) = \frac{1}{\Gamma(\nu-\alpha)} \frac{d^\nu}{dt^\nu} \int_0^t (t-t')^{\nu-\alpha-1} f(t') dt' \tag{C3}$$

In particular, for $0 < \alpha < 1$, so that $n=1$, we get

$$_0D_t^\alpha f(t) = \frac{1}{\Gamma(1-\alpha)} \frac{d}{dt} \int_0^t (t-t')^{-\alpha} f(t') dt' = \frac{1}{\Gamma(-\alpha)} \int_0^t (t-t')^{-(\alpha+1)} f(t') dt' \tag{C3'}$$

The name of fractional derivative given to the operator $_0D_t^\alpha$ follows from the fact that the fractional derivative of order α of an arbitrary power of a variable – and then the fractional derivative of any function which may be expanded in a power series - is performed with the same rules of the usual derivative of integer order, viz.,

$$[_0D_t^\alpha t^\nu] \equiv \frac{d^\alpha}{dt^\alpha} t^\nu = \frac{\Gamma(\nu+1)}{\Gamma(\nu-\alpha+1)} t^{\nu-\alpha} \tag{C4}$$

In particular, for $\nu=0$, we get a result at first sight surprising: the fractional derivative of a constant is a function, namely

$$\left[ _0D_t^\alpha 1 \right] \equiv \frac{d^\alpha}{dt^\alpha} 1 = \frac{1}{\Gamma(1-\alpha)} t^{-\alpha} \tag{C4'}$$

However, note that, if α is positive and integer, the fractional derivative is zero as it must be, due to the divergence of the gamma function. It is worthwhile mentioning that in the above case the fractional derivative as a function of α oscillates around the zero values taken in correspondence of the integer values of α.

Let us now come back to the inversion of eq. (15). Multiplying the *lhs* by the denominator of the *rhs* yields

$$c_\alpha \hat{\hat{P}}(k,u) + \sigma^2 \tau^{-\alpha} k^2 \left[ u^{-\alpha} \hat{\hat{P}}(k,u) \right] - c_1 \tau^{1-\alpha} \left[ u^{1-\alpha} \hat{\hat{P}}(k,u) \right] = -c_1 \tau^{1-\alpha} u^{-\alpha} + \frac{c_\alpha}{u}$$

The inverse L-transform $(u \to t)$ is readily obtained from eq. (C2) for the factors in square parentheses and from usual algebra for the others

$$c_\alpha \hat{P}(k,t) + \sigma^2 \tau^{-\alpha} k^2 \,_0D_t^{-\alpha} \hat{P}(k,t) - c_1 \tau^{1-\alpha} \,_0D_t^{1-\alpha} \hat{P}(k,t) = -\frac{c_1}{\Gamma(\alpha)} \left( \frac{\tau}{t} \right)^{1-\alpha} + c_\alpha$$

Inverse F-transforming and then differentiating with respect to time yields eq. (16).